# Temperature and Electron Concentration Dependences of 1/$f$ Noise in Hg$_{1-x}$Cd$_x$Te - Evidence for a Mobility Fluctuations Mechanism

Adil Rehman,[*a] Volodymyr Petriakov,[a] Ivan Yahniuk,[a,b] Aleksandr Kazakov,[c] Iwona Rogalska,[d] Jakub Grendysa,[a,d] Michał Marchewka,[d] Maciej Haras,[e,f] Tomasz Wojtowicz,[c] Grzegorz Cywiński,[a,e] Wojciech Knap,[a,e] and Sergey Rumyantsev,[*a,e]

[a.] *CENTERA Laboratories, Institute of High Pressure Physics, Polish Academy of Sciences, Warsaw 01-142, Poland.*

[b.] *Terahertz Center, University of Regensburg, 93040 Regensburg, Germany.*

[c.] *International Research Centre MagTop, Institute of Physics, Polish Academy of Sciences, Warsaw 02-668, Poland.*

[d.] *University of Rzeszów, Institute of Materials Engineering, Center for Microelectronics and Nanotechnology, 35-959 Rzeszów, Poland.*

[e.] *Centre for Advanced Materials and Technologies CEZAMAT, Warsaw University of Technology, Warsaw 02-822, Poland.*

[f.] *Gdańsk University of Technology, Faculty of Electronics, Telecommunications and Informatics, Advanced Materials Center, 80-233 Gdańsk, Poland*.

* E-mail: adilrehhman@gmail.com; roumis4@gmail.com

**Abstract**

Hg$_{1-x}$Cd$_x$Te is a unique material with the band-gap tunable by the temperature, pressure, and cadmium content in a wide range, from 1.6 eV to inverted band-gap of -0.3 eV. This makes Hg$_{1-}$ $_x$Cd$_x$Te one of the key materials for infrared and terahertz detectors, whose characteristics largely depend on the material noise properties. In this work, we investigated the low-frequency 1/$f$ noise in a thick (800 nm) HgCdTe layer and in a field effect transistor (FET) with an 8 nm wide HgTe quantum well. Both structures exhibited a small contribution from contact noise and showed weak noise dependences on temperature. Investigation of the 1/$f$ noise in HgTe quantum well FET as a function of gate voltage revealed that the noise also depends weakly on electron concentration. These findings indicate that the noise properties of Hg$_{1-x}$Cd$_x$Te are similar to those of graphene, where mobility fluctuations were found to be the dominant mechanism of the 1/$f$ noise.

**Introduction**

Mercury cadmium telluride (Hg$_{1-x}$Cd$_x$Te) has been known as a narrow-band semiconductor for over 60 years (see ref.[1] and references therein). It is one of the key materials for infrared (IR) and far-infrared (FIR) detectors and IR imaging systems.[1-4] It also attracts significant attention for terahertz applications.[5-9] The band structure of Hg$_{1-x}$Cd$_x$Te strongly depends on temperature (*T*), pressure, and cadmium content (x), which allows the bandgap tuning from 1.6 eV to an inverted bandgap of -0.3 eV.[1] At a critical value of x = 0.12 and *T* = 300 K, the separation between the valence and conduction bands at the Γ point of the Brillouin zone becomes zero, and the dispersion law of electrons and light holes becomes linear. This offers great opportunities for the design of photodetectors with controlled properties and enables the exploration of interesting phenomena arising from the linear dispersion of bands, as in graphene, but in a bulk material.

The key characteristic of IR detectors is their detectivity, which is always limited by some form of noise. Thermal and shot noises are important for the detectors operating at high modulation frequencies. For the sensitive detectors operating at low modulation frequency, the low-frequency $1/f$ and generation-recombination noise play a vital role. Therefore, many publications are devoted to studying the low-frequency noise in HgCdTe photodetectors, mainly in photodiodes.[10-17] These studies are of great practical importance, but it is difficult to analyse the noise mechanism in devices of complex design, especially with the potential barriers.

To understand the nature of noise in a given material, it is important to investigate the simple ohmic structures. Studying the nature of noise in the $Hg_{1-x}Cd_xTe$ material itself is important for understanding the $1/f$ noise mechanisms in narrow-band semiconductors and for the design of infrared and terahertz detectors, quantum computers, and other advanced applications where noise control is critical. However, there are only a few publications, where the $1/f$ noise was studied in just HgCdTe material [18-21] or in HgTe quantum well (QW).[22]

Here, we studied the electrical and low-frequency $1/f$ noise properties of bulk and QW $Hg_{1-x}Cd_xTe$ structures. Our findings show that the noise properties of $Hg_{1-x}Cd_xTe$ are unusual and resemble those of graphene. The temperature dependence of the noise in a bulk HgCdTe layer and the dependence of the noise on electron concentration in HgTe QW provide a strong argument for mobility fluctuations as the dominant mechanism of the $1/f$ noise in $Hg_{1-x}Cd_xTe$. These findings contradict previous studies that attributed the noise to carrier number fluctuations in HgCdTe-based systems.[18-22]

**Experimental Methods**

In this work, two different types of $Hg_{1-x}Cd_x Te$-based devices were studied. The first one represents the 800 nm thick layer of $Hg_{0.79}Cd_{0.21}Te$, which we refer to as bulk samples. (see Fig. 1a). The second type of the device was the HgCdTe/HgTe heterostructure. The investigated structure had a field effect transistor (FET) geometry (see schematic view in Fig. 1(b)), which allowed the tuning of carrier concentration in the HgTe QW channel by the gate voltage ($V_G$).

The bulk $Hg_{1-x}Cd_x Te$ structure was grown on polished CdZnTe (211) substrate by molecular beam epitaxy (MBE) in the Riber Compact 21 MBE system equipped with a liquid Hg source and reflection high-energy electron diffraction (RHEED) system. The growth rate and composition (x = 0.21) of the $Hg_{1-x}Cd_x Te$ epilayers were controlled and the process was monitored in a real-time via in situ RHEED system (see ref.[23] for detailed process). The grown stack consisted of a 50 nm $Hg_{0.3}Cd_{0.7}Te$ cap layer and an active 800 nm thick $Hg_{0.79}Cd_{0.21}Te$ layer on CdZnTe substrate (see Fig. 1a). To fabricate the Ohmic contacts, the windows in the cap layer were first opened using standard photolithography and wet etching in the 0.05% bromine-methanol solution. Afterward, indium (In) contacts were deposited via electrodeposition technique. The final device configurations were obtained in a second photolithography step and wet mesa etching to identify and separate each device.

The HgCdTe/HgTe heterostructure were grown by MBE on a (013)-oriented GaAs substrates.[4] The 8 nm QW was encapsulated in HgCdTe barrier layers with high molar Cd content (see Fig. 1b for the grown stack details). Electron beam lithography and wet etching in the 0.08% bromine-methanol solution were used to define 5×3 $\mu m^2$ devices, and indium contacts, acting as a source and drain terminals were made. The plasma-enhanced chemical vapor deposition technique was employed to deposit a 140 nm thick $Si_3N_4$ layer as a top gate dielectric. Titanium/gold (Ti/Au) metal layers (10/110 nm) acting as a top gate contact was thermally evaporated. A detailed description of the device fabrication process can be found elsewhere.[24]

All measurements were conducted by placing the devices on the sample stage of a closed cycle cryogenic probe station (Lake Shore Inc., CRX-VF),[25] and special probes were used for the electrical and low-frequency noise measurements as a function of temperature without the need of the probes lifting. Electrical measurements were performed by using a semiconductor parametric analyzer (Keithley 4200A-SCS). For the noise studies, the voltage fluctuations across the load resistor ($R_L$) were amplified by a low-frequency noise amplifier and analyzed by a fast Fourier transform "PHOTON" dynamic signal analyzer. The obtained spectral noise density ($S_V$) was then converted into the spectral noise density of short circuit current fluctuations ($S_I$) as $S_I = S_V[(R_L + R_D)/(R_L R_D)]^2$. Here, $R_D$ is the device resistance. The background noise of the system was always 15-20 dB smaller than that of the studied devices and was checked by replacing the device with a metal resistor of the same value as $R_D$.

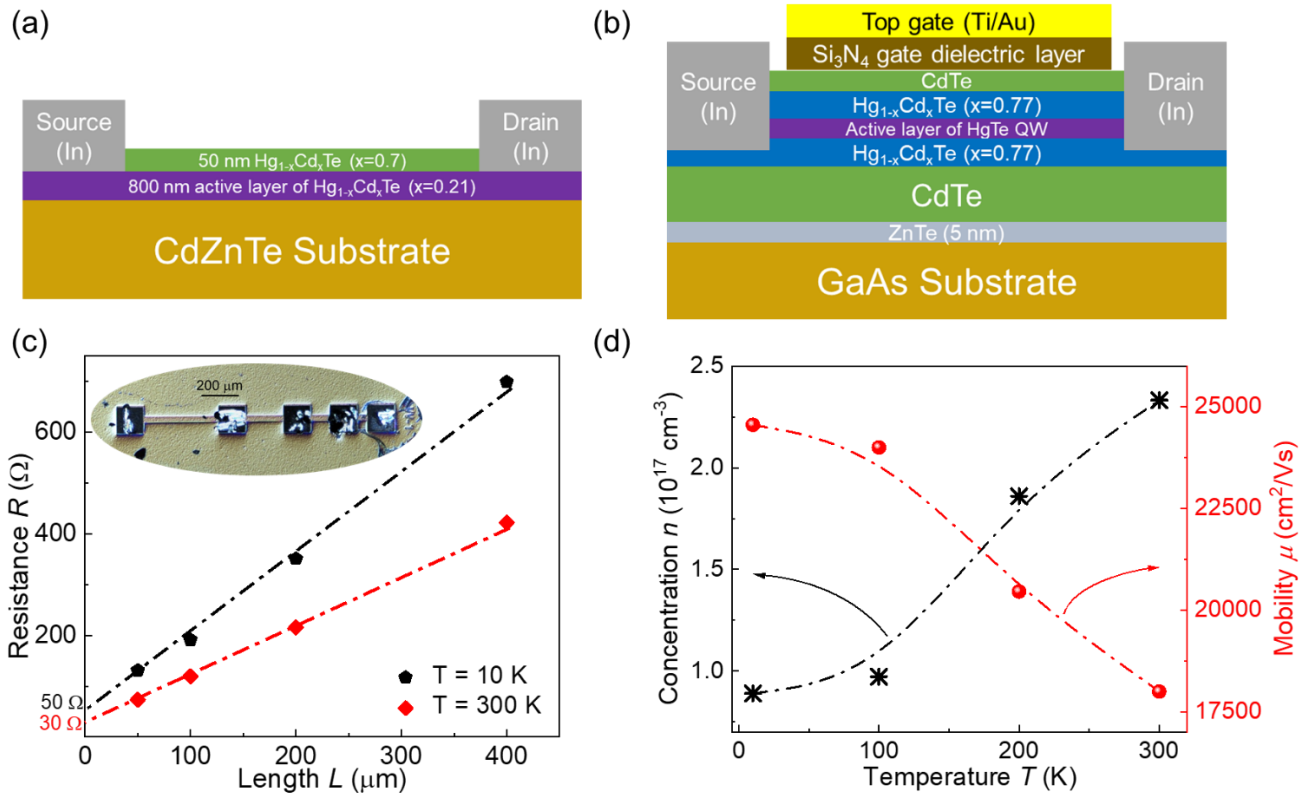

**Fig. 1.** Cross-sectional schematic views of the (a) bulk HgCdTe and (b) HgTe QW devices. (c) Resistance as a function of bulk HgCdTe device length at $T$ = 10 K and $T$ = 300 K. The solid symbols correspond to the individual devices of different lengths, and dash-dot lines are a guide for the eyes. The inset shows the optical image of the studied TLM structure. (d) Charge carrier concentration and mobility of a bulk HgCdTe device at different temperatures.

**Results and discussion**

The bulk HgCdTe devices had the shape of the 30 µm wide stripes of various lengths (see inset in Fig. 1c). This allowed us to estimate the total contact resistance ($2R_c$) and contact noise using a transmission line model (TLM) approach.[26, 27] The current-voltage (I-V) characteristics of all devices were linear within a voltage range from zero to 150 mV. Fig. 1(c) shows the resistance as a function of device length (*L*) for one of the TLM structures at two different temperatures. As seen, the dependences of resistances on the device length are linear. The intercept with the Y-axis yields the total contact resistance. It can be seen in Fig. 1(c) that at both low and high temperatures, the total contact resistance is several times smaller than the sample resistance. The Hall bars geometry was also included in the bulk HgCdTe structure to estimate the mobility and charge carrier concentrations. Fig. 1(d) shows the measured concentration (*n*) and mobility ($\mu$) of electrons at different temperatures. It can be seen that as temperature increases, the concentration increases and mobility decreases.

Figure 2(a) shows examples of the noise spectra at 10 K for several bulk HgCdTe devices of different lengths. The noise spectra of all devices had the $1/f^\gamma$ shape with $\gamma$ = 1.1 - 1.3. Figure 2(b) shows the spectral noise density of short circuit current fluctuations ($S_I$) as a function of current ($I$) for the devices of different lengths at $f$ = 2 Hz. It can be seen that $S_I$ is proportional to the square of the current (i.e. $S_I \propto I^2$), which implies that the resistance fluctuations are responsible for the $1/f$ noise and current only makes these fluctuations visible.

One of the ways to distinguish between contact and channel noise is to examine the noise as a function of the device length. The spectral noise density of current fluctuations can be written as:

$$\frac{S_I}{I^2} = \frac{S_{R_{ch}}}{R_{ch}^2} \frac{R_{ch}^2}{(R_c+R_{ch})^2} + \frac{S_{R_c}}{R_c^2} \frac{R_c^2}{(R_c+R_{ch})^2} \tag{1}$$

Here, $R_{ch}$ is the channel resistance, and ($S_{R_{ch}}/R_{ch}^2$) and ($S_R/R_c^2$) are the relative spectral noise densities of the channel and contacts resistance fluctuations, respectively. If $R_{ch} \gg R_c$ and the noise from the channel dominates, the spectral noise density of current fluctuations can be written as $(S_I/I^2) = (S_{R_{ch}}/R_{ch}^2) \propto L^{-1}$. On the other hand, If noise originates from the contacts, the spectral noise density of current fluctuations can be written as $(S_I/I^2) = (S_{R_c}/R_c^2)(R_c^2/R_{ch}^2) \propto L^{-2}$.

Fig. 2(c) shows how the noise depends on the device length at 10 K. Red and blue dash-dot lines represent the shape of the dependences calculated based on the first and second terms of Eq. (1) with the actual values of the contact and channel resistances. As seen, the experimental dependence of noise on the device length confirms that noise originated from the channel, not from the contacts.

The inset of Fig. 2(c) shows the resistance of one of the investigated bulk HgCdTe devices as a function of temperature. The solid symbols represent the experimental data points, and the dash-dot line shows the resistance calculated using the mobility and concentration obtained from the Hall measurements. As seen, the resistance of the bulk HgCdTe device decreases significantly with increasing temperature and coincides well with the calculated red dash-dot line. Figure 2(d) shows the temperature dependence of the noise in bulk HgCdTe devices. As seen, although the resistance changes with temperature, the noise ($S_I/I^2$) exhibits weak temperature dependence. It was also found that noise spectra do not change significantly with the temperature.

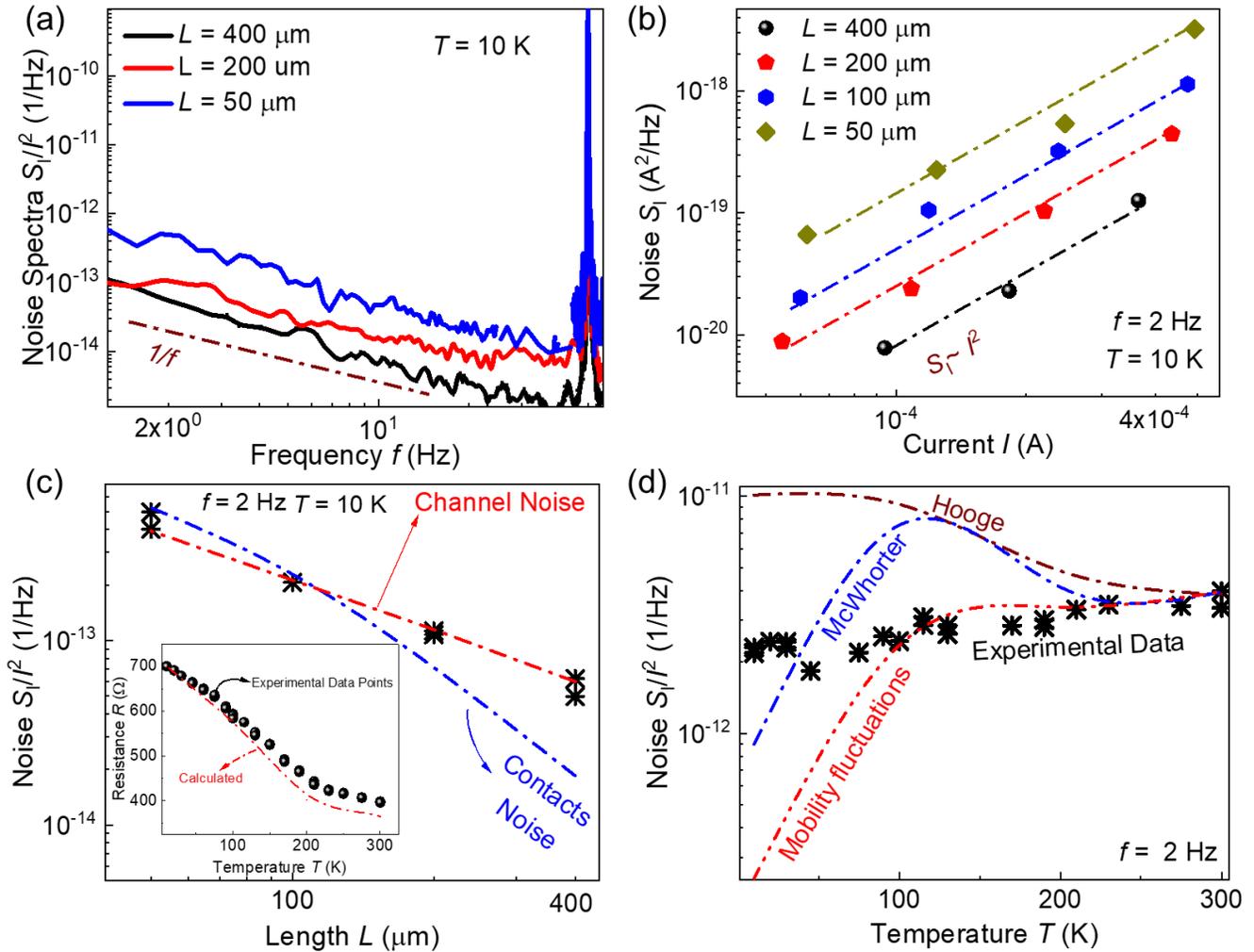

**Fig. 2.** (a) Noise spectra of bulk HgCdTe devices of different lengths at 10 K. (b) Spectral noise density of current fluctuation as a function of current for different bulk HgCdTe devices. (c) Noise as a function of length for a HgCdTe TLM structure. Red and blue dash-dot lines represent the shape of the noise dependences in the case of the noise from channel and contacts, respectively. Inset shows the resistance of one of the investigated devices as a function of temperature. The red dash-dot line shows the resistance calculated using the concentration and mobility of charge carriers obtained from the Hall measurements. (d) Noise as a function of temperature for one of the investigated HgCdTe devices. Dash-dot lines show the hypothetical behavior of noise calculated assuming the Hooge formula, McWhorter number of carriers fluctuations model, and mobility fluctuations model.

We can analyse the shape of the temperature dependence of the noise based on known models. The Hooge's empirical formula claims that noise is of the volume origin and is inversely proportional to the total number of carriers (N):[28]

$$\frac{S_I}{I^2} = \frac{\alpha_H}{fN}, \qquad (2)$$

where $\alpha_H$ is the constant. Another option is that noise is due to the tunnelling of electrons to the adjusting $Hg_{0.3}Cd_{0.7}Te$ layer or/and CdZnTe substrate. This mechanism is described by the McWhorter's model:[29]

$$\frac{S_I}{I^2} = \frac{kTN_t}{\alpha f W L n_s^2}, \qquad (3)$$

where $k$ is the Boltzmann constant, $T$ is the temperature, $N_t$ is the effective trap density, $WL$ is the channel area, $n_s$ is the two-dimensional concentration, and $\alpha$ is the attenuation coefficient of the electron wave function under the barrier.

In general, the 1/$f$ noise can also originate from the traps in the bulk. However, simple estimates show that this narrow-band semiconductor at a doping level of more than $10^{17}$ cm$^{-3}$ is degenerate at all studied temperatures (*i.e.* the Fermi level is in the conduction band). This implies that all traps are filled, and do not contribute to noise. Therefore, we assume that this mechanism cannot make a significant contribution to noise.

Another possible origin of the noise is the mobility fluctuations. This mechanism of noise was discussed in refs.[30, 31] In accordance with the model, this kind of noise is due to the fluctuations of the scattering cross section ($\sigma$). The spectral noise density of current fluctuations for one type of scattering centers can be written as:

$$\frac{S_I}{I^2} = \frac{4N_{t\mu}\tau P(1-P)}{1+(\omega\tau)^2} l_0^2 (\sigma_2 - \sigma_1)^2. \qquad (4)$$

Here $N_{tu}$, $\tau$, $P$, $l_o$, and $V$ represent the scattering centers' concentration, characteristic time constant, probability of scattering centers to be in a state with a cross-section $\sigma_1$, carriers mean free path, and volume, respectively.

It can be seen from Eq. (2) that Hooge's formula predicts that the noise ($S_I/I^2$) should depend on temperature as $1/N$. McWhorter's model predicts that noise depends on temperature as $\propto T/n_s^2$ for uniform energy distribution of traps. To obtain the $1/f$ noise due to the mobility fluctuations one needs to integrate Eq. (4) over multiple scattering centers with a wide distribution of characteristic times ($\tau$). Although the result depends on the actual distribution of the scattering centers, as a first approximation, we can assume that noise linearly increases with the temperature increase (see ref.[31] for the detailed analysis). Since mobility is proportional to the mean free path, we can write the spectral noise density of current fluctuations in the case of mobility fluctuation mechanism as:

$$\frac{S_I}{I^2} \sim T\mu^2 \qquad (5)$$

The red, blue, and brown dash-dot lines in Fig. 2(d) are calculated based on the three models discussed above. The noise values are fitted to the experimental data at $T$ = 300 K. As can be seen, the mobility fluctuations model fits the experimental data very well at high temperatures. At low temperatures, none of the models is consistent with the experimental data. However, it is important to mention that this is a more general problem. The majority of the models predict that the noise decreases and tends to zero with temperature approaching zero (one of the exceptions is the Hooge's formula). However, this is rarely observed experimentally in any electronic system. Therefore, we conclude that the mobility fluctuations is the most realistic mechanism of the 1/$f$ noise in HgCdTe.

We also investigated the electrical and low-frequency noise characteristics of the HgTe QW structures. Fig. 3(a) shows the dependence of the HgTe QW-based device resistance on the gate voltage at two temperatures. It can be seen that the device behaves as an n-channel FET. For the gate voltages $V_G$< -1.5V, the resistance increases, manifesting a decrease in the electron concentration. While for the gate voltage $V_G$>-1.5V, the resistance only weakly depends on gate voltage and temperature indicating the dominance of the contact resistance. The inset in Fig. 3a shows the gate voltage dependence of the electron concentration in the QW calculated as $n_s = C(V_G - V_t)/q$, where $C$ is the gate capacitance per unit area, $q$ is the elementary charge, and $V_t$ is the threshold voltage taken to be $V_G$=-4.5 V.

The dependence of spectral noise density of short circuit current fluctuations (at $f$=10 Hz) as a function of current at $V_G$= -4 V and $T$ = 100 K is shown in the inset in Fig. 3(b). It can be seen that $S_I$ is proportional to the square of the current. Since channel resistance dominates at $V_G$ = -4 V (see Fig. 3a), it implies, as described earlier, that channel resistance fluctuations are responsible for the 1/$f$ noise origin. The noise spectra of the studied device at different temperatures are shown in Fig. 3(b). It can be seen that the noise spectra have the 1/$f$ shape, and, as with the bulk HgCdTe device, neither the noise amplitude nor the shape of the spectrum changes significantly with temperature.

Figure 3(c) shows the resistance and noise of HgTe QW-based FET as a function of temperature at $V_G$= -4 V, where the device resistance is dominated by the channel resistance. The bandgap of HgTe QW strongly depends on temperature.[32, 33] The detailed band structure calculations based on the eight-band k·p Hamiltonian for (013)-oriented heterostructures, which directly takes into account the interactions between $\Gamma_6$, $\Gamma_8$, and $\Gamma_7$ bands in bulk materials were performed in ref.[34] The calculations show that at $T$ = 10 K the band structure is inverted, and with the temperature increase, the band gap decreases and becomes zero at $T \approx 100$ K. It can be seen that resistance decreases as temperature increases, but the noise, despite the significant temperature dependence of the band gap,[34] shows only a weak temperature dependence. This behaviour is similar to that observed for a bulk HgCdTe device (see Fig. 2(d)).

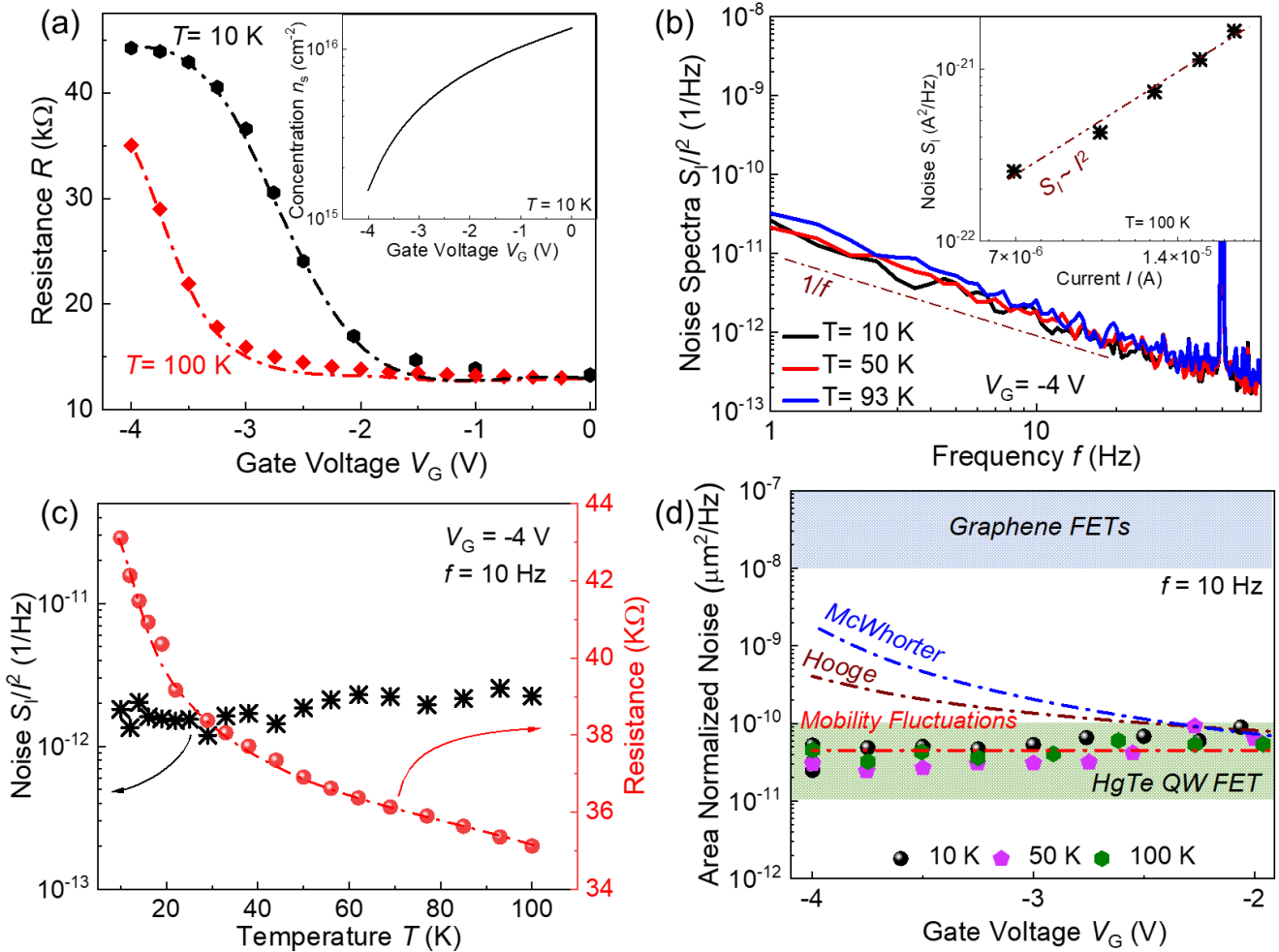

**Fig. 3.** (a) Resistance of a HgTe QW-based FET as a function of gate voltages at two temperatures. (b) Noise spectra at different temperatures for a HgTe QW-based FET. Inset shows the spectral noise density of current fluctuation as a function of current at 100 K. (c) Resistance and noise as a function of temperature at $V_G$= -4 V. (d) Symbols show the area normalized noise as a function of gate voltage at three different temperatures. The upper dashed area corresponds to the noise level in graphene devices.[32] The red, blue, and brown dash-dot lines show the hypothetical behavior of the gate voltage dependences of noise assuming the Hooge formula, McWhorter number of carriers fluctuations model, and mobility fluctuations model.

Figure 3(d) shows the area normalized noise (Area x $S_I/I^2$) of the studied HgTe QW-based FET (at $f$= 10 Hz) as a function of $V_G$ at three different temperatures. The noise ($S_I/I^2$) is normalized to the sample area to facilitate comparison with graphene devices (light blue area in Fig. 3d). The typical noise range for graphene devices is in the range of $10^{-8}$–$10^{-7}$ µm²/Hz.[35] It is seen that the noise level in the studied HgTe QW is smaller than in graphene and does not depend on $V_G$.

The dash-dot lines in Fig. 3(d) show the shapes of the gate voltage dependences of noise assuming Hooge's formula, McWhorter's number of carriers fluctuations, and mobility fluctuations models assuming mobility independent of gate voltage (concentration). It is important to emphasize that, in contrast to the number of carrier's fluctuation mechanism of noise, the contribution to noise from the mobility fluctuations does not depend on either the total number of carriers or their concentration. Therefore, only the mobility fluctuations model is consistent with the observed weak dependence of noise on the gate voltage. This means that the noise properties of $Hg_{1-x}Cd_xTe$ are similar to those of graphene, where the mobility fluctuations mechanism is considered as the dominant source of the $1/f$ noise.[36]

**Conclusions**

The low-frequency $1/f$ noise in bulk HgCdTe structure and HgTe QW was investigated over a wide temperature range. Both structures exhibited weak temperature dependence of the noise. A study of the 1/$f$ noise in FETs based on HgTe QW showed that the noise was also weakly dependent on electron concentration. These results indicate the similarity of the noise properties of $Hg_{1-x}Cd_xTe$ and graphene, provide strong arguments for the mobility fluctuations mechanism of the 1/$f$ noise in $Hg_{1-x}Cd_xTe$, and open the pathways to optimize the performance of HgCdTe devices for infrared and terahertz detection applications.

**Author contributions**

S.R. conceived the idea of study. A.R. performed the measurements and analyzed the data. I.Y, V.P, M.H, and A.K fabricated the devices. I.R, J.G, and M.M. grow the HgCdTe wafer. W.K, G.C, and T.W contributed to the data analysis. S.R. and A.R. wrote the manuscript. All authors participated in the manuscript preparation.

**Conflicts of interest**

There are no conflicts to declare.

**Data availability**

The data supporting the findings of the study are available within the article. Additional data that support the findings of this study are available from the corresponding author upon reasonable request

**Acknowledgements**

The work was supported by the European Union (ERC "TERAPLASM", project number: 101053716). Views and opinions expressed are however those of the authors only and do not necessarily reflect those of the European Union or the European Research Council. Neither the European Union nor the granting authority can be held responsible for them. This work was also partially supported by the "Center for Terahertz Research and Applications (CENTERA2)" project (FENG.02.01-IP.05-T004/23) and by the "MagTop" project (FENG.02.01-IP.05-0028/23) both carried out within the "International Research Agendas" programme of the Foundation for Polish Science co-financed by the European Union under the European Funds for a Smart Economy 2021-2027 (FENG). The authors are also very thankful to N. N. Mikhailov, S. A. Dvoretsky for providing HgCdTe/HgTe QW structures.